\documentclass{article}
\usepackage[T1]{fontenc} 
\usepackage[utf8]{inputenc} 
\usepackage{ismir,amsmath,cite,url,amssymb}
\usepackage{graphicx}
\usepackage{color}
\usepackage[normalem]{ulem}
\newcommand{\mulan}{MuLan}
\usepackage{lineno}

\title{MuLan: A Joint Embedding of Music Audio and Natural Language}

\iftrue
\multauthor
{Qingqing Huang$^1$ \hspace{1cm} Aren Jansen$^1$ \hspace{1cm} Joonseok Lee$^{1,2}$} 
{\bfseries{Ravi Ganti$^1$ \hspace{1cm} Judith Yue Li$^1$ \hspace{1cm} Daniel P. W. Ellis$^1$}\\
Google Research$^1$, Seoul National University$^2$\\ 
{\tt\small \{qqhuang, arenjansen, joonseok, gmravi, judithyueli, dpwe\}@google.com}}

\def\authorname{Q. Huang, A. Jansen, J. Lee, R. Ganti, J. Y. Li, and D. P. W. Ellis}
\fi

\begin{document}

\maketitle
\begin{abstract}
Music tagging and content-based retrieval systems have traditionally been constructed using pre-defined ontologies covering a rigid set of music attributes or text queries. This paper presents \mulan{}: a first attempt at a new generation of acoustic models that link music audio directly to unconstrained natural language music descriptions. \mulan{} takes the form of a two-tower, joint audio-text embedding model trained using 44 million music recordings (370K hours) and weakly-associated, free-form text annotations. 
Through its compatibility with a wide range of music genres and text styles (including conventional music tags), the resulting audio-text representation subsumes existing ontologies while graduating to true zero-shot functionalities.  
We demonstrate the versatility of the \mulan{} embeddings with a range of experiments including transfer learning, zero-shot music tagging, language understanding in the music domain, and cross-modal retrieval applications. 
\end{abstract}
\section{Introduction}\label{sec:introduction}

Classifiers are generally trained to label examples with predefined and fixed class inventories, which are often manually specified as a structured ontology indicating inter-class relationships.  Empowered by recent advances in neural language modeling and their demonstrated transfer learning competence, researchers have begun exploring less restrictive natural language interfaces to access the categorical information underlying raw content signals.  The majority of this work has been in the visual and audio event domain, where a recent series of studies have demonstrated the utility of jointly embedding media content with natural language captions~\cite{jia2021scaling,radford2021learning,koepke2022audio,kilgour2022text,nagrani2022learning}.  These joint embeddings have demonstrated strong capabilities in a range of applications, including transfer learning, cross-modal retrieval, automatic captioning, and zero-shot classification.  

The success of these efforts strongly depends on large-scale training resources and hefty neural network architectures that are flexible enough to model the complex, non-monotonic relationship between language and other modalities.  
In particular, the visual domain has greatly benefited from the availability of large amounts of captioned images available across the web~\cite{jia2021scaling}.  However, in the general environmental audio domain, such large-scale audio-caption pairs are less readily available and related efforts have relied on small captioned datasets~\cite{drossos2020clotho,kim2019audiocaps}.  Critically, these datasets do not span the diversity of sound-descriptive language and their success in the more difficult zero-shot setting has been lacking~\cite{favory2020coala,favory2021learning,koepke2022audio}.

This paper considers this task of jointly embedding audio and natural language, but focuses specifically on the music domain.  Our goal is to produce a flexible language interface with which any musical concept can be linked to related music audio.  We face similar training data prerequisites to works listed above.
However, while general environmental audio consists of background sounds that are unlikely to elicit unprompted description, music audio is often a central focus.  Consequently, text associated with music videos is much more likely to relate to the underlying musical concepts that we aim to model (e.g., genres, artists, moods, structure). 
Thus, our strategy is to assemble a collection of textual annotations extracted from metadata, comments, and playlist data and map them to a training set of over 44 million internet music videos.
As was the case with image-text model training in~\cite{jia2021scaling}, our text data only truly refers to the musical content in a fraction of cases. Therefore, we also explore text pre-filtering using a text classifier separately trained to identify music descriptions.

We use this large-scale dataset to train \mulan{}, a new generation of semantically-structured music audio embedding model equipped with a natural language interface.
\mulan{} employs a two-tower parallel encoder architecture, using a contrastive loss objective that elicits a shared embedding space between music audio and text.  
We demonstrate that \mulan{} not only leads to state-of-the-art performance in transfer learning for various music information retrieval tasks, but also enables a range of functionalities in cross-modal text-to-music retrieval, zero-shot music tagging, and music-domain language understanding.

\section{Related Work}
\label{sec:related}

\noindent\textbf{Audio representation learning.}
Transfer learning using large-scale, task-agnostic pretraining of general-purpose content representations has become a dominant approach in several fields.
Audio representation learning has been no exception, including both 
general environmental audio~\cite{gong2021ast,gong2021ssast} and music audio~\cite{hamel2013transfer,van2014transfer,choi2017transfer,dhariwal2020jukebox}.  
Different pretraining mechanisms have been explored.
In supervised pretraining, an Audio Spectrogram Transformer (AST)~\cite{gong2021ast}, pretrained on ImageNet~\cite{deng2009imagenet} and AudioSet~\cite{gemmeke2017audio}, achieved state-of-the-art results in various tagging tasks. A strong early baseline for music audio representation learning was provided in \cite{van2014transfer}, using the Million Song Database~\cite{bertin2011million}.

In unsupervised and self-supervised pretraining, both discriminative and generative model approaches have been demonstrated to be successful. 
Discriminative training was explored in \cite{jansen2018unsupervised, saeed2021contrastive, turpault2019semi, spijkervet2021contrastive} where the models tried to learn representations that assign higher similarity to audio segments extracted from the same recording compared to segments from different recordings. SSAST~\cite{gong2021ssast} explored similar discriminative losses, as well as generative masked spectrogram patch modeling.  It was shown in~\cite{castellon2021codified} that the intermediate embedding of a generative model also provides a strong audio representation for downstream classification.
Various forms of weak supervision, such as user interaction statistics and visual cues, have also been examined in ~\cite{oramas2017multi, huang2020large,ferraro2021enriched}. 

Our work focuses on developing a recipe of cross-modal supervision using an abundance of text annotations that are weakly associated with the music audio.
We benchmark the transfer learning capabilities of the learned representations against analogous past work, and also evaluate different audio encoder architectures.  

\noindent\textbf{Cross-modal contrastive learning.}
Spurred by the success of using contrastive learning to align image features and free-form natural language using large-scale data~\cite{jia2021scaling,radford2021learning}, tri-modal architectures were proposed in~\cite{guzhov2022audioclip} and~\cite{nagrani2022learning} where an audio tower was introduced to the image-text model and contrastive learning is used to enforce the cross-modal alignment.
Along the same line in the audio domain,
\cite{favory2020coala} used contrastive learning to align the latent representation of audio and associated tags. The tags come from a fixed vocabulary of size 1K from Freesound~\cite{font2013freesound}, and the input to the text encoder was the multi-hot encoded tags.
Follow up work in~\cite{favory2021learning} uses a pretrained, non-contextual word embedding (Word2Vec) model to support generalization to new terms beyond the 1K tags.  However, this still does not support generalization to free-form natural language. 
Contrastive learning was also explored in \cite{xie2021zero} for zero-shot audio classification, using AudioSet and ESC-50\cite{piczak2015esc} data. 
Our method focuses on mining a much larger scale collection of audio-text pairs specifically for the music domain.  Our data scale supports using state-of-the-art Transformer-based audio and contextual language encoders, which 
led to a truly arbitrary zero-shot music tagging and retrieval for the first time.

\noindent\textbf{Music text joint embedding models.}
Content-based music information retrieval requires linking the rich semantics expressible to free-form text with both broad and fine-grained musical properties.
One approach is to consider a large number of text label classes and try to ground the semantics in music with a multi-label classification task. 
In \cite{huang2020large}, a large vocabulary of 100K $n$-grams was mined from noisy natural language text associated with music videos.  Then, a cross entropy loss was employed to train the music audio encoder, where the softmax layer weights served as text label embeddings that were aligned with audio features by construction. 
The work in \cite{won2021emotion} explored various training tasks (classification, regression, metric learning) to align free-form text and music audio, relying on pre-existing emotion labels to connect the modalities. 

Closest to our work is MuLaP\cite{manco2022learning}, where 250K audio-caption pairs were mined from a private production music library and used to train a multimodal Transformer with early fusion of the two modalities. 
Their choice of early fusion, as accomplished with cross-attention layers, restricts the utility of the resulting embeddings to transfer learning applications.  Critically, our two-tower parallel encoder approach results in a joint embedding space that provides a natural language interface to arbitrary music audio.  This opens up downstream opportunities for cross-modal retrieval, zero-shot tagging, and language understanding.

\begin{figure}
  \centering \includegraphics[width=1.0\columnwidth]{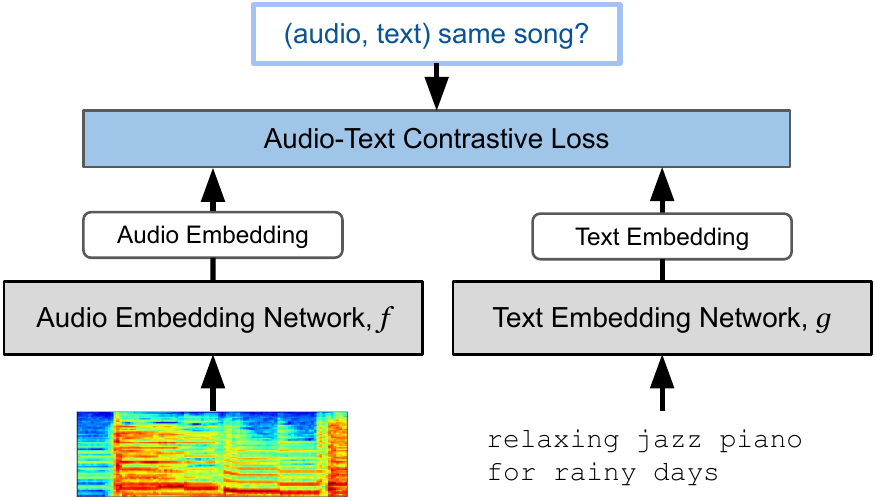}
  \caption{Learning framework diagram.}
  \label{fig:schem}
\end{figure}

\section{Proposed approach}\label{sec:proposed}

Our goal is to construct a shared embedding space for music audio and free-form natural language text, in which proximity is predictive of shared semantics both within and across modalities.  To accomplish this, we rely on cross-modal contrastive learning and a simple two-tower architecture.  This is a highly data-intensive endeavor, which we support by mining a large-scale training dataset of (audio, text) pairs.
We describe these components in turn below.

\subsection{Learning Framework}
\label{sec:proposed-framework}
Figure~\ref{fig:schem} shows a high-level schematic of the learning framework.  Each \mulan{} model consists of two separate embedding networks for the audio and text input modalities.  These networks share no weights, but each terminates in $\ell_2$-normalized embedding spaces with the same dimensionality, $d$.  The audio embedding network, $f: \mathbb{R}^{F \times T} \rightarrow \mathbb{R}^d$, takes as input log mel spectrogram context windows with $F$ mel channels and $T$ frames.  The text embedding network, $g: \mathcal{A}^n \rightarrow \mathbb{R}^d$ takes as input a null-padded text token sequence of length $n$ over a token vocabulary $\mathcal{A}$. 

Given a set of music recordings and the associated text elements for each recording, we construct a cross-modal training dataset of  (audio, text) pairs as follows. For each recording, we compute an $F$-channel log mel spectrogram and extract a collection of $T$-frame context windows.  We null-pad or truncate each associated text element to a fixed length $n$.  Then, each mini-batch $\mathcal{B}$ consists of a set of $B$ target audio-text pairs of the form $\{(\mathbf{x}^{(i)}, \mathbf{t}^{(i)})\}_{i=1}^B$.  Here, each target pair is sampled by first selecting a random recording and sample a random spectrogram context window ${\mathbf{x}^{(i)}} \in \mathbb{R}^{F \times T}$ from it. Next, we randomly select one of its associated text elements ${\mathbf{t}^{(i)}} \in \mathcal{A}^n$.
%
%
This sampling scheme means that multiple epochs are required to cover the entirety of the training audio and all the associated text.
We also experimented with concatenating multiple text annotations for each example, but it did not generally work as well.

We train to minimize a batch-wise Contrastive Multiview Coding loss function~\cite{tian2020contrastive}, which is a cross-modal extension of the popular InfoNCE and NT-Xent losses~\cite{oord2018representation,chen2020simple}.  
For each batch $\mathcal{B}$, this loss {$\mathcal{L} (\mathcal{B})$} takes the form
\begin{align*}
    \sum_{i=1}^B - \log \left[ 
    \frac{h[f({\mathbf{x}^{(i)}}), g({\mathbf{t}^{(i)}})]}
    {\sum\limits_{j \ne i} h[f({\mathbf{x}^{(i)}}), g({\mathbf{t}^{(j)}})] + h[f({\mathbf{x}^{(j)}}), g({\mathbf{t}^{(i)}})]} \right],
  \label{eq:loss}
\end{align*}
where $h$ is a critic function given by $h[\mathbf{a},\mathbf{b}] = \exp(\mathbf{a}^T\mathbf{b} / \tau)$ for $\mathbf{a},\mathbf{b} \in \mathbb{R}^d$, and $\tau \in (0, 1]$ is a trainable temperature hyperparameter. For our $\ell_2$-normalized embedding model outputs, the inner product is effectively cosine similarity. The critic's goal is to produce a large positive value for target audio-text pairs, and a small value close to zero for all non-target pairs constructed within the batch. 
Temperature values less than one function to increase the output range of $h$.  
Previous research\cite{lee2020large,chen2020simple} demonstrated that a large batch size is beneficial 
to contrastive loss optimization.

\subsection{Audio Embedding Network}
\label{sec:proposed-audio}
For the audio embedding tower, $f$, we consider two proven audio architectures.  Following its introduction to the audio machine learning community~\cite{hershey2017cnn}, the \emph{Resnet-50} architecture has become a common and well-performing option.  It is a straightforward adaptation of the original vision architecture: as in~\cite{hershey2017cnn}, we remove the stride of 2 in the first convolutional layer and apply to log mel spectrograms ($F\!=\!64$ mel channels, 25 ms Hanning window, 10 ms step size) treated as grayscale images.  Unlike the Resnet-50 model in~\cite{hershey2017cnn} which operated on 0.96-second context windows, in order to allow the modeling of longer-term musical structure, our implementation takes as input 10-second windows (randomly selected from each training clip), in the form of $(F\!=\!64) \times (T\!=\!1000)$ spectrogram patches.
During training, we apply SpecAugment to each spectrogram using the parameters from~\cite{gong2021ast} before passing it into the embedding network.  A final mean pooling operation is applied across time and mel channels, followed by a linear fully connected layer with $d=128$ units, whose output is $\ell_2$-normalized.  We pretrain all but the final linear transform layer via logistic regression on AudioSet~\cite{gemmeke2017audio}, including all 527 classes, and discard the final classifier layer before fine-tuning for our task.

\emph{Audio Spectrogram Transformer} (AST) is a port of the successful Vision Transformer (ViT) base architecture and is currently the state-of-the-art in the audio event classification space~\cite{gong2021ast}.  AST consists of a stack of 12 Transformer blocks (hidden dimension 768, 12 self-attention heads) that are applied to a sequence of ``tokens'' corresponding to a flattened set of linear-transformed $16 \times 16$ (stride 10 along both axes) time-frequency patches extracted from the $(F \!=\!128) \times (T\!=\!1000)$ log mel spectrogram context windows.  We again apply SpecAugment during training. Similar to the Transformer-based language models, trainable positional encodings are added to the sequence of patch tokens, and a \texttt{[CLS]} token is prepended to the sequence as a summary of the contextual patch embeddings.
We apply a linear fully-connected layer with $d=128$ units and $\ell_2$-normalization to the final 768-dimensional encoding at the \texttt{[CLS]} token position, and this forms the output of audio embedding network $f$. We warm-start training for all but the final linear transform layer using the public AST checkpoint~\cite{gong2021ast}.

\subsection{Text Embedding Network}
\label{sec:proposed-text}
For the text embedding model, we consider the commonly-used Bidirectional Encoder Transformer (BERT) with base-uncased architecture~\cite{devlin2018bert}, which consists of a stack of 12 Transformer blocks (hidden dimension of 768 and 12 self-attention heads). We apply the BERT wordpiece tokenizer to convert a text input string into a sequence of tokens ($n=512$). The output of the text embedding network is defined to be the \texttt{[CLS]} token embedding, linearly transformed to the shared audio-text embedding space of dimension $d=128$ and subsequently $\ell_2$-normalized. We warm-start our text embedding network using the publicly available checkpoint~\cite{BertBaseUncasedCkpt}. 

\subsection{Training Dataset Mining}
\label{sec:data}
To assemble a large-scale collection of (audio, text) pairs needed to train our \mulan{} embedding models, we start with a collection of 50 million internet music videos. From the soundtrack of each video, we extract a 30-second clip starting at the 30 second mark.  We then apply a pre-existing music audio detector and discard any clip that is less than half music content. After this filtering, we are left with approximately 44 million 30-second clips, which amounts to nearly 370K hours of audio. 

\begin{table}[t!]
    \vspace{-0.2cm}
    \caption{Text annotation examples.}
    \vspace{0.1cm}
    \label{tab:text-label-ex}
    \centering
    \footnotesize{
    \begin{tabular}{l|l}
    \hline
        {\textbf{Type}} & {\bf Examples}
        \\
        \hline
         Short-form (SF) & tags like genre, mood, instrument, artist name, \\
         {} & song title, album name \\
         \hline
         Long-form  (LF) & `Hip-hop features rap with an electronic backing.' \\
         {} & `The melody is so nostalgic and unforgettable.' \\
         \hline
         Playlist   (PL) & `Feel-good mandopop indie', `Latin workout' \\
         {}         & `Salsa for broken hearts', `Piano for study' \\
         \hline
    \end{tabular}
    }
    \vspace{-0.3cm}
\end{table}

\begin{table}[t!]
\vspace{-0.2cm}
\caption{Statistics for text data sources. Tokens counts (in billions) are across all 44M videos.  APV is the average number of text annotations (i.e. separate free-form strings) per video, including those with none.}
\vspace{0.1cm}
  \label{tab:textdata}
  \centering \small
  \begin{tabular}{l|cc|cc}
    \hline
               & \multicolumn{2}{c|}{\bf Pre-filter}   & \multicolumn{2}{c}{\bf Post-filter} \\
    \hline           
    {\bf Type}    & Tokens (B) & APV  &  Tokens (B) &  APV  \\
    \hline
    Short-form     & 31.2 & 42.9 & 5.4  & 29.6 \\
    Long-form      & 30.7 & 70.7 & 0.2  & 0.4  \\
    Playlists      &  2.5 & 24.3 & -    & -    \\
    \hline
  \end{tabular}
\end{table}

For each music video, we consider 3 sources of noisy text data: (i) \emph{short-form} (SF) text including video titles and tags; (ii) \emph{long-form} (LF) text including video descriptions and comments; and (iii) titles of 171 million \emph{playlists} (PL) that are linked to the internet music videos in our dataset.
None of these text sources is guaranteed to be referring to the musical properties of the soundtrack. In particular, comments data contains the most noise, and can be subjective or less directly related to the music content.
In Table~\ref{tab:text-label-ex}, we show examples that are indeed music-related to give the readers a flavor of each type of text annotation. 

In observance of the highly noisy text, we experimented with training \mulan{} with the SF and LF text data filtered to a cleaner set of music-descriptive annotations (PL is used unfiltered). 
For this, we fine-tune a pre-trained BERT model with a binary classification task on a small curated set of 700 sentences, which are manually labeled to be music-descriptive or not. We then apply this text classifier to filter the sentences in the LF annotations.
Separately, we apply a set of rule-based filtering heuristics to clean up the SF annotations. 
Table~\ref{tab:textdata} shows the size and coverage of each of these text sources, both before and after filtering. Note that playlist titles and filtered long-form annotations are only available for a minority of recordings in the dataset (18M and 6.8M out of the total 44M, respectively).

We also convert AudioSet into a set of audio-text pairs, denoted below as ASET.  Specifically, we include all examples for all 527 classes, using each label string attached to an example as an associated text annotation.  This results in a set of approximately 2 million 10-second clips for training, each with 1.8 label annotations on average.
Given the great scale imbalance of these four different data sources, which is often at odds with their linguistic richness and quality, we construct each mini-batch with a prescribed set of proportions that were chosen without any optimization: 2:2:1:1 for SF:LF:PL:ASET.  This means that despite its small scale, the (e.g.) filtered LF annotations still comprise 1/3 of each mini-batch.

\section{Experiments}\label{sec:experiments}

We evaluate MuLan using both the Resnet-50 audio encoder (M-Resnet-50) and AST audio encoder (M-AST). In both cases we use the BERT-base-uncased architecture as the text encoder. 
We train all models for 14 epochs on the collection of audio-text pairs mined from the 44M music recordings and the processed text labels in all categories: AudioSet (ASET), short-form tags (SF), long-form sentences (LF), playlist information (PL). We use the Adam optimizer 
with a step decay learning rate schedule using a decay factor 0.9 applied every 40K steps and initial values of $5\!\times\! 10^{-5}$ for M-Resnet-50 and $4\!\times\! 10^{-5}$ for M-AST. The temperature parameter is initialized to $\tau=0.1$ for all models.  M-Resnet-50 is trained with a batch size of $B=6144$ pairs, while $B=5120$ pairs were used for M-AST due to memory limitations.
Since M-AST and M-Resnet-50 show roughly similar performance in the evaluation tasks considered, we use M-Resnet-50 throughout the text ablation study for its better training efficiency.

\subsection{Evaluation Tasks}\label{sec:eval}

\subsubsection{Zero-shot Music Tagging}
Given a music clip and a set of candidate text label tags, we define each prediction score as the cosine similarity between the audio embedding of the music clip and the text embedding of each tag string.
The generalization ability of the proposed method to potentially unseen target labels is achieved through (i) the use of a contextual text encoder, which provides a flexible prediction space, and (ii) the use of cross-modal contrastive learning to anchor the language semantics to an audio representation. 
 
We conduct this evaluation with two music tagging benchmarks: MagnaTagATune (MTAT)~\cite{law2009input} and the music related portion of AudioSet~\cite{gemmeke2017audio}.
For MagnaTagATune, we consider both the well-exercised top-50 tag set, as well as the full 188 tag set. We use standard train/validation/test partitions (note that zero-shot experiments do not use train/validation) and report class-balanced area under the receiver operating characteristic curve (AUC-ROC) on the test set. 
The audio clips in MagnaTagATune are 29 seconds long, so we split each into 3 non-overlapping 10-second segments and average the segment-level embeddings to get the clip-level embedding.
For AudioSet, we consider a 25-way genre tagging task (Gen-25) as studied in~\cite{huang2020large}, and a richer 141-way tagging task (Mu-141) that includes the entire music subtree of AudioSet ontology.

It is important to note that AudioSet is included in contrastive training, and a fraction of MTAT classes overlap with the AudioSet ontology. As a result, AudioSet and (to lesser extent) MTAT evaluations are not strictly zero-shot from a label exposure perspective.  However, the explicit, matched AudioSet supervision is diluted by the abundance of free-form language supervision during \mulan{} training.  Therefore, by comparing \mulan{} models and conventional AudioSet classifiers, we can measure the cost of moving to a flexible natural language interface that additionally supports classes outside the AudioSet ontology.  

\subsubsection{Transfer Learning with Linear Probes}
In addition to the zero-shot experiments introduced above, we also evaluate the audio encoder as a general purpose feature extractor for downstream tagging tasks. We again consider the two benchmarks of MagnaTagATune and AudioSet, and use the training datasets to train an independent per-class logistic regression layer on top of the frozen 128-dimensional audio embeddings.
We follow the same evaluation protocol of past transfer learning studies using these datasets, allowing for a direct comparison of performance.

\subsubsection{Music Retrieval from Text Queries}
\label{sec:music-retrieval-eval}
Given a music search collection and a text query, \mulan{} provides the ability to retrieve the music clips that are closest to the query in the embedding space.
This evaluation is relevant to music retrieval applications, where content features can offer finer-grained and more complete similarity information when compared with metadata-based methods~\cite{typke2005survey}. 
We consider a proprietary collection of 7,000 expert-curated playlists, which do not overlap with the playlist information used in training. Each expert-curated playlist has a title and a description, and consists of 10-100 music recordings. The playlist titles are usually short phrases, including a mixture of genres, sub-genres, moods, activities, artist names, and compositional elements (e.g. \emph{`Indie Pop Workout'}, \emph{`Relaxing Korean Pop'}). Playlist descriptions consist of one or more complete sentences (see pos/neg entries of "Playlist" row of Table~\ref{tab:text-triplet-eval} for examples).
The playlist evaluation includes approximately 100K unique recordings.

We construct two cross-modal retrieval evaluation sets from the expert-curated playlist data, one using titles as queries and the other using descriptions.  For each dataset, we use the recordings belonging to the corresponding playlist as the ground truth retrieval targets, and all the 100K recordings as the pool of candidates. We report both AUC-ROC and mean average precision (mAP).
We use the same embedding averaging and cosine similarity-based scoring mechanism as in the zero-shot tagging case. However, the playlist information is of substantially different nature compared to the tags involved in the music tagging benchmarks. Instead of a small vocabulary of mostly basic genres and instruments, the playlist titles and descriptions have much finer-grained information and are similar to 
queries that are presented to music search engines.
 
\begin{table}[t!]
    \vspace{-0.2cm}
    \caption{Text triplet evaluation examples.}
    \vspace{0.1cm}
    \label{tab:text-triplet-eval}
    \centering
    \small{
    \begin{tabular}{p{0.14\columnwidth}|p{0.74\columnwidth}}
    \hline
        {\textbf{Eval Set}} & {\bf Anchor / Positive / Negative}
        \\
        \hline
         Ontology &  Steelpan / Sounds of a tuned percussion instrument originally constructed from steel oil drums by hammering out small patches on the head to produce separate pitches. / The sound of a musical instrument that produces sound by vibration of air in a tubular resonator in sympathy with the vibration of the player's lips. \\
         \hline
         Playlist & Relaxing Korean Pop / Lets make your chill mood with a collection of easy-going sounds from Korean artists. / These fun and upbeat songs from the alternative side of the pop music spectrum will keep you energized while you exercise. \\
         \hline
    \end{tabular}
    }
\end{table}

\begin{table}
\vspace{-0.2cm}
\caption{Music tagging results reported in AUC-ROC.}
\vspace{0.1cm}
  \label{table:zeroshot}
   \small
   \centering{
  \begin{tabular}{l|c @{\hspace{1.5\tabcolsep}} c|c @{\hspace{1.5\tabcolsep}} c}
    \hline
    {} & \multicolumn{2}{c|}{\bf AudioSet} & \multicolumn{2}{c}{\bf MTAT} \\
    {\bf Model} & {\bf Gen-25} & {\bf Mu-141} & {\bf Top-50} & {\bf All-188} \\
    \hline
    \multicolumn{5}{l}{{\bf (a) Zero-shot} (Trained w/ ASET + SF + LF + PL)} \\
    \hline 
        M-AST          & 0.840    & 0.909    & 0.778    & \bf{0.776}   \\
        M-Resnet-50    & 0.840    & 0.899    & \bf{0.782}    & 0.772   \\
    \hline
    \hline
    \multicolumn{4}{l}{{\bf (b) Text ablation} (using M-Resnet-50 Zero-shot)} \\
    \hline
    \hspace{0.1cm}ASET + SF + LF       & 0.839    & 0.907    & 0.760    & 0.756   \\
    \hspace{0.1cm}ASET + SF          & 0.839    & 0.885    & 0.754    & 0.747   \\
    \hspace{0.1cm}ASET             & \bf{0.886}    & \bf{0.942} & 0.753  & 0.771   \\
    \hspace{0.1cm}SF/LF Unfiltered    & 0.845    & 0.908    & 0.774    & 0.766   \\
    \hline
    \hline
    \multicolumn{4}{l}{{\bf (c) Linear probe}} \\
    \hline
    M-AST              & 0.906    & \bf{0.942}    & 0.925    & 0.953   \\ 
    M-Resnet-50        & \bf{0.910}    & 0.940    & \bf{0.927}    & \bf{0.954}   \\
    \hline
    {\emph{Baselines:}} & & \\
    \hspace{0.1cm}Hybrid~\cite{huang2020large}
                    & 0.904    & 0.920    & 0.915    & 0.941   
    \\
    
    \hspace{0.1cm}JukeBox~\cite{dhariwal2020jukebox,castellon2021codified}   & -    & -    & \hspace{4pt}0.915{$^*$}    & -   \\
    \hspace{0.1cm}MuLaP~\cite{manco2022learning}      & -    & -    & \hspace{4pt}0.893{$^*$}    & -   \\
    \hspace{0.1cm}CLMR~\cite{spijkervet2021contrastive} & -    & -    & \hspace{4pt}0.866{$^*$}    & -   \\
    \hline
    \hline
    \multicolumn{4}{l}{{\bf (d) End-to-end training baselines}} \\ 
    \hline
    \hspace{0.1cm}AST~\cite{gong2021ast}
                    & 0.888    & \bf{0.949}    & -    & -   \\
    \hspace{0.1cm}SC-CNN~\cite{won2020evaluation}
                    &               - & -                    & 0.913{$^*$}    & -   \\
    \hline
    
    \hline
  \end{tabular}
  }
  \scriptsize
  \noindent $^*$ indicates that the number is brought from the original paper. \\
\vspace{-0.1cm}
\end{table}

\subsubsection{Text Triplet Classification}
\label{sec:text-triplet-eval}
Compared to the conventional pre-trained BERT model, our text encoder is fine-tuned using in-domain music data and cross-modal contrastive loss.  Note that there are no text-only training objectives. To measure whether our proposed method deepens the text encoder's understanding of music related text, we directly evaluate the text embeddings with a triplet classification task. Each triplet consists of 3 text strings of the form of (\textit{anchor}, \textit{pos}, \textit{neg}), and it is considered correct if \textit{pos} is closer than \textit{neg} to \textit{anchor} in the text embedding space.
We derive two such text triplet evaluation sets.  The first uses the AudioSet ontology~\cite{gemmeke2017audio}: for each of the 141 music related classes, we use its label string as the anchor text, its long-form description as the positive text, and sample 5 random class's long-form description as the negative text to construct 5 triplets.
For the second set, we sample 1,000 triplets from the expert-curated playlist data in a similar fashion: we first sample a playlist, set the anchor and positive text to be its title and description, respectively, and then set the negative text to be the description of another randomly sampled playlist. Examples of both sets are shown in Table~\ref{tab:text-triplet-eval}.

\subsection{Results and Discussion}

\subsubsection{Music Tagging}

Table~\ref{table:zeroshot}(a) shows the zero-shot tagging metrics, where M-Resnet-50  and M-AST obtain comparable performance. 
Note that there can be a significant misalignment between the word sense of a label in the tagging evaluation compared to that in our training text.  This can lead to a degradation in performance relative to the explicitly supervised linear probe setting where the task-expected tag semantics can be learned.  The MTAT gap is substantially larger than AudioSet's, driven by particularly bad performance for (i) MTAT tags with nonspecific meaning or multiple senses, e.g. ``weird'' and ``beats''; and (ii) MTAT tags involving simple negation (e.g. ``not rock'', ``no piano''). This is a result of the text encoder not adequarely modeling the meaning of these negated concepts, which is a well known problem with BERT~\cite{ettinger2020bert,tejada2021study} (the text embedding of ``not rock'' is similar to ``rock'' and performance suffers).
 
Table~\ref{table:zeroshot}(b) shows the results of the text ablation study, which aims to understand the benefits of different sources of text labels. 
Note that as we remove each dataset we maintain the same proportions described in Section~\ref{sec:data}.
Unsurprisingly, training with AudioSet alone gets the highest AUC in AudioSet evaluation, with the text encoder learning the exact label semantics reflected in the test data.  On the other hand, including more data sources in general improves performance on all other downstream tasks (MTAT, retrieval/text triplet evaluations in Tables~\ref{table:query} and~\ref{table:text-triplet}) and the loss on AudioSet AUC is relatively minor.
%
We observe that for the music tagging tasks considered, training with unfiltered data actually achieves comparable performance compared to the filtered version.  That the model learns similarly useful associations without being overwhelmed by the sheer amount of noise in the raw text data came as a surprise.  We speculate that our text filtering was too aggressive, having removed annotations that were not obviously music-related, but semantically important nonetheless.  Since contrastive learning is highly noise tolerant, the gain from restricting to more strongly aligned audio-text pairs may have been offset by the loss of a large set of additional useful pairs.  

\begin{table}[t!]
\vspace{-0.2cm}
\caption{Text query music retrieval evaluation results. Text ablation/unfiltered models use M-Resnet-50.}
\vspace{0.1cm}
  \label{table:query}
  \centering \small
  \begin{tabular}{l|cc|cc}
    \hline
    {}  & \multicolumn{2}{c|}{\bf Title}   & \multicolumn{2}{c}{\bf Description} \\
    {\bf Model} &   AUC   &   mAP                &   AUC   &   mAP  \\
    \hline
    M-AST        & \bf{0.933}  & 0.110   & \bf{0.903}   & \bf{0.090}    \\
    M-Resnet-50  & 0.931       & 0.104        & 0.901        & 0.084   \\
    \hline
    {\emph{Text Ablation:}} & & \\
    \hspace{0.2cm}ASET+SF+LF     & 0.917  & 0.101   & 0.892    & 0.077   \\
    \hspace{0.2cm}ASET+SF         & 0.913  & 0.089   & 0.867    & 0.060   \\
    \hspace{0.2cm}ASET             & 0.626  & 0.005   & 0.688    & 0.009   \\
    \hline
    \emph{SF/LF Unfiltered}   & \bf{0.933} & \bf{0.111} & 0.897    & 0.081   \\
    \hline
  \end{tabular}
  \vspace{-0.4cm}
\end{table}

\begin{table}[t!]
\caption{Text triplet classification accuracy AudioSet ontology evaluation and Playlist title to description evaluation. Text ablation/unfiltered models use M-Resnet-50.}
\vspace{0.1cm}
  \label{table:text-triplet} \small
  \centering
  \begin{tabular}{l|c|c}
    \hline
    {\bf Model} & {\bf Playlist}  &  {\bf AudioSet} \\
    \hline
    M-AST         & \bf{0.959} & \bf{0.962} \\
    M-Resnet-50   & 0.945 & 0.951 \\
    \hline
    {\emph{Text Ablation:}} & & \\
    \hspace{0.2cm}ASET + SF + LF      & 0.935 & 0.952 \\
    \hspace{0.2cm}ASET + SF         & 0.910 & 0.938\\
    \hspace{0.2cm}ASET            & 0.693 & 0.818\\
    \hline
    \emph{SF/LF Unfiltered}    & 0.949  & 0.959 \\
    \hline
    {\emph{Baselines:}} & & \\
    \hspace{0.2cm}SimCSE~\cite{gao2021simcse}  & \bf{0.950}  & 0.938\\
    \hspace{0.2cm}SBERT~\cite{reimers2019sentence}  & 0.942  & 0.889\\
    \hspace{0.2cm}USE~\cite{cer2018universal}   & 0.918   & \bf{0.946}\\ 
    \hspace{0.2cm}BERT~\cite{devlin2018bert}  & 0.850  & 0.847 \\
    \hline
  \end{tabular}
\end{table}

Table~\ref{table:zeroshot}(c) shows that when applying linear probes on MuLan audio embeddings, we achieve SOTA transfer learning performance on all tagging tasks.  This demonstrates that MuLan's pretrained audio encoder continues to produce high quality general-purpose music audio embeddings, while also supporting new natural language applications.
Finally, Table~\ref{table:zeroshot}(d) lists end-to-end training baselines for 3 of these tasks.  Our linear probe results exceed 2 of 3, and only slightly trails a SOTA AST AudioSet classifier.

\subsubsection{Music Retrieval from Text Queries}
In Table~\ref{table:query}, we evaluate MuLan models (including with text/filter ablation) on the query retrieval evaluation tasks
introduced in Section~\ref{sec:music-retrieval-eval}. 
Even though we start with a BERT checkpoint pretrained with massive language resources, training MuLan with only AudioSet clips and label annotations provides very limited ability to ground in-domain natural language to music. Such limited cross-modal supervision does not generalize to the rich semantics that appear in the playlist titles and descriptions, which are more in line with the complex queries that are presented to real-world music search engines. 
We observe significant gain after including the large-scale short-form tags mined from the internet, which helps the model learn to ground more fine-grained music concepts. 
There is additional gain when including comments and playlist data, where the complete sentences are helpful for grounding the more complex queries, including multi-term queries (e.g.\emph{`instrumental action movie soundtrack'}), compositional queries (e.g. \emph{`classical music with middle eastern influence'}), and even queries with negation (e.g. \emph{`hard rock without vocals'}).
Again, we find that training is surprisingly robust to annotation noise, achieving similar performance using unfiltered training text.

\subsubsection{Text Triplet Classification}

Table~\ref{table:text-triplet} lists triplet classification accuracy on evaluations 
introduced in Section~\ref{sec:text-triplet-eval}. 
We compare MuLan text embedding against the following baselines: Sentence Transformer~\cite{reimers2019sentence}, SimCSE
~\cite{gao2021simcse}, Universal Sentence Embedding~\cite{cer2018universal}, and the average token embedding of BERT-base-uncased (this outperforms the \texttt{[CLS]} encoding by a large margin). All baselines are Transformer-based models with similar size to ours. The first three were trained with sentence-level contrastive loss, while BERT is trained with masked language prediction.  We warmstart the MuLan text encoder using this same BERT baseline, 
but it is subsequently only trained with the cross-modal loss.
We find that when including our long-form text annotations, the resulting text embedding model, which is now specialized to the music domain, outperforms the generic sentence embedding models.  While it is not surprising in-domain text is helpful, it is remarkable that successful specialization is accomplished without using any text-only fine-tuning loss.

\vspace{-0.1cm}
\section{Conclusions}\label{sec:conclusion}
We presented a music audio and natural language joint embedding model trained with an unprecedented scale of weakly paired text and audio data. Our experiments demonstrate the versatility of the natural language interface in a range of applications.
The pretrained audio embeddings also achieve SOTA transfer learning performance on music tagging benchmarks.  This is a first attempt at building a free-form natural language interface for music audio and there is plenty of room for improvement.  Specifically, we believe improved text filtering methods that better distinguish weak signal from absolute noise will result in better handling of rare and subtle language constructs. 

\clearpage

\bibliography{main}

\end{document}